\documentstyle[aps,prl,psfig,multicol]{revtex}

\tighten
\begin{document}
\preprint{MPI H-V17-1997}  
\bibliographystyle{prsty}
\def\graphpath{}
\def\bibpath{}
\title{On the Role of the Annihilation Channel in Front Form Positronium}

\author{Uwe Trittmann}

\address{Max-Planck-Institut f\"ur Kernphysik \\
D-69029 Heidelberg  \\
}
\date{\today}


\maketitle

\begin{abstract}
The annihilation channel is implemented into the {\em front form}
calculations of the positronium spectrum presented in  
Ref.~\cite{Trittmann97a}. The effective Hamiltonian
is calculated analytically. Its eigensolutions are obtained numerically.
A complete separation of the dynamical and instantaneous part of the
annihilation interaction is observed.
We find the remarkable effect that the annihilation channel stabilizes
the cutoff behavior of the spectrum. 

\end{abstract}

\begin{multicols}{2}
\narrowtext


\section{Introduction}

Recently, a calculation of the positronium spectrum within the {\em front form}
formalism was presented 
\cite{Trittmann97a}. In this work, the entire spectrum was calculated,
{\em i.e.}~all sectors defined by the kinetic component of the 
total angular momentum, $J_z$, were considered. However, the
one photon annihilation channel
was omitted. It is the purpose of the present paper to include this
channel into the calculations, which are completed with this step. 

The article is structured as follows. In Section \ref{Section:AnniChannel}
the implications of introducing the annihilation channel into the positronium
theory are stated. Next, the spectrum of the ``full'' positronium is 
calculated and the results are presented. Section \ref{Section:ParamDep}
deals with the parameter dependence of the results, {\em i.e.}~the dependence 
on cutoffs and convergence parameters. A discussion of the results follows.
The calculation of the matrix elements of the effective Hamiltonian 
is reviewed in Appx.~\ref{Appx:MatrixElements}.

\section{The annihilation channel in positronium}
\label{Section:AnniChannel}

It is a topic of its own merit to construct a manageable Hamiltonian 
for the positronium system.
One possible way was shown in Ref.~\cite{Trittmann97a}, 
where an effective  Fock space, consisting of two sectors
$|e\bar{e}\rangle$ and $|e\bar{e}\gamma\rangle$, named $P$- and $Q$-space 
was considered. One can justify the absence of any higher 
Fock state \cite{CommentHigherStates}
from the structure of the applied
formalism of effective interactions \cite{Pauli96b}. 
The general formalism is set up for a 
non-abelian $SU(N)$ gauge theory. Unlike QED, the one boson
state is absent in these theories because of  color neutrality. 
Nevertheless, one has to take care of the
one photon state $|\gamma\rangle$ in QED, and in what follows, we demonstrate
how this can be done.

Firstly, it is important to notice not only the differences between 
the QED Table \ref{HolyMatrixQED} and the analogous QCD table in 
\cite{Pauli96b}, but also
the similarities. 
Some of the graphs occuring in QCD are absent in QED. In particular
those graphs with a three- or four-boson interaction and 
the instantaneous interactions connecting four bosons. 
But, although an additional sector occurs as a first row and a first 
column in the 
QED table,
neither is a change in the higher Fock sectors observed, nor is the ordering 
altered in any way.

\noindent
\begin{minipage}{8.6cm}
\begin{table}[t]
\def\d{$\bullet$}  \def\v{ V }  \def\b{ $\cdot$ } \def\s{ S }   \def\f{ F }
\begin {tabular}{|l|r||cc|ccc|cccc|ccccc|}
\hline 
  \rule[-3mm]{0mm}{8mm}  Sector & $n$ & 
     0 & 1 & 2 & 3 & 4 & 5 & 6 & 7 & 8 & 9 &10 &11 &12 &13 
\\ \hline \hline   
    $ |\gamma \rangle$ &  0 & 
    \d & \v &\b &\f &\b &\b &\b &\b &\b &\b &\b &\b &\b &\b 
\\
    $ |e\bar e\rangle $ &  1 & 
    \v &\d &\s &\v &\f &\b &\f &\b &\b &\b &\b &\b &\b &\b 
\\ \hline
    $ |\gamma\ \gamma\rangle$ &  2 & 
    \b & \s &\d &\v &\b &\v &\f &\b &\b &\b &\b &\b &\b &\b 
\\
    $ |e\bar e\, \ \gamma \rangle$ &  3 & 
    \f &\v &\v &\d &\v &\s &\v &\f &\b &\b &\b &\b &\b &\b 
\\
    $ |e\bar e\, e\bar e\rangle $ &  4 & 
    \b&\f &\b &\v &\d &\b &\s &\v &\f &\b &\b &\f &\b &\b 
\\ \hline
    $ |\gamma \ \gamma \ \gamma \rangle$ &  5 & 
    \b& \b &\v &\s &\b &\d &\v &\b &\b &\v &\f &\b &\b &\b 
\\
    $ |e\bar e\, \ \gamma \ \gamma\rangle $ &  6 & 
    \b&\f &\f &\v &\s &\v &\d &\v &\b &\s &\v &\f &\b &\b 
\\
    $ |e\bar e\, e\bar e\, \ \gamma \rangle$ &  7 & 
    \b &\b &\b &\f &\v &\b &\v &\d &\v &\b &\s &\v &\f &\b 
\\
    $ |e\bar e\, e\bar e\, e\bar e\rangle $ &  8 & 
    \b&\b &\b &\b &\f &\b &\b &\v &\d &\b &\b &\s &\v &\f 
\\ \hline
    $ |\gamma \ \gamma \ \gamma \ \gamma \rangle$ & 9 & 
    \b&\b &\b &\b &\b &\v &\s &\b &\b &\d &\v &\b &\b &\b 
\\
     $ |e\bar e\, \ \gamma \ \gamma \ \gamma \rangle$ & 10 & 
    \b&\b &\b &\b &\b &\f &\v &\s &\b &\v &\d &\v &\b &\b 
\\
    $ |e\bar e\, e\bar e\, \ \gamma \ \gamma \rangle$ & 11 & 
    \b&\b &\b &\b &\f &\b &\f &\v &\s &\b &\v &\d &\v &\b 
\\
     $ |e\bar e\, e\bar e\, e\bar e\, \ \gamma \rangle$ & 12 & 
    \b&\b &\b &\b &\b &\b &\b &\f &\v &\b &\b &\v &\d &\v 
\\
     $ |e\bar e\, e\bar e\, e\bar e\, e\bar e\rangle$ & 13 & 
    \b&\b &\b &\b &\b &\b &\b &\b &\f &\b &\b &\b &\v &\d 
\\ \hline
\end{tabular}
\vspace{0.3cm}
\caption{\label{HolyMatrixQED}The 
Hamiltonian matrix for QED. 
The sectors $n$ are numbered starting at zero.
The vertex,
seagull and fork interactions are denoted by V, S, F respectively. Diagonal
matrix elements are symbolized by \d . 
}
\end{table}
\end{minipage}
In addition to the $P$- and $Q$-spaces defined above, we introduce the
$N$-space, {\em  i.e.~}the sector containing the $|\gamma\rangle$ states, 
into the system. The corresponding projector is
\begin{equation}
\hat{N} := \sum_{n}^{\rm all\;\; QN} 
|(\gamma)_n\rangle \langle (\gamma)_n|.
\end{equation}
The whole procedure of subsequent projections of higher Fock states onto the 
remaining Hilbert space of states can be carried out like before 
\cite{Trittmann97a} until 
one arrives at a $(2\times 2)$ matrix, 
operating in the $N$- and $P$-space rather than in the $P$- and $Q$-space
\begin{equation}\label{NPSpace}
H_{\rm LC}\; \psi= \left(
                \begin{array}{cc}
                        H_{NN} & H_{NP}\\
                        H_{PN} & H_{PP}\\
                \end{array}
             \right) 
                \left(
                \begin{array}{c}
                        \psi_{\gamma}\\
                        \psi_{e\bar{e}}\\
                \end{array}
             \right) =\omega
                \left(
                \begin{array}{c}
                        \psi_{\gamma}\\
                        \psi_{e\bar{e}}\\
                \end{array}
             \right). 
\end{equation}
From Table \ref{HolyMatrixQED} one can read off the 
interaction of the one photon
state with all other sectors: the vertex interaction annihilates the photon
into an electron-positron pair and a fork interaction scatters it into the 
sector $|e\bar{e}\gamma\rangle$. The latter interaction is already contained
in the effective interaction, Eq.~(\ref{NPSpace}), because of the projection
of the $Q$-space. 


\begin{figure}[t]
\centerline{
\psfig{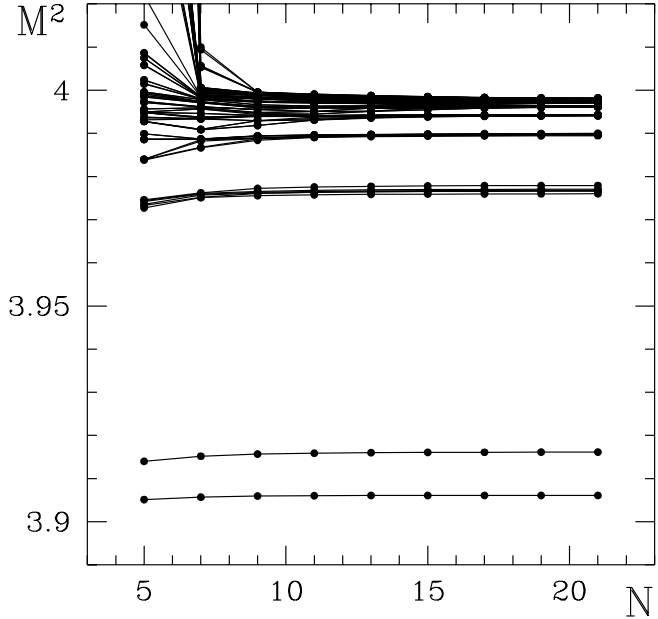}}
\vspace{0.3cm}
\caption{\label{specanni}
The positronium spectrum including the annihilation channel in the 
$J_z{=}0$ sector. Parameters of the calculation: $\alpha=0.3, 
\Lambda=1.0\, m_f$. 
The mass squared eigenvalues $M^2_n$ in units of the electron mass $m^2_f$
are shown as functions of the number of integration points $N\equiv N_1=N_2$.
The triplet states, especially $1^3S_1$, are lifted up by the annihilation 
interaction, whereas the singlet
mass eigenvalues stay the same.}
\end{figure}
\noindent
\begin{minipage}{8.6cm}
\begin{figure}
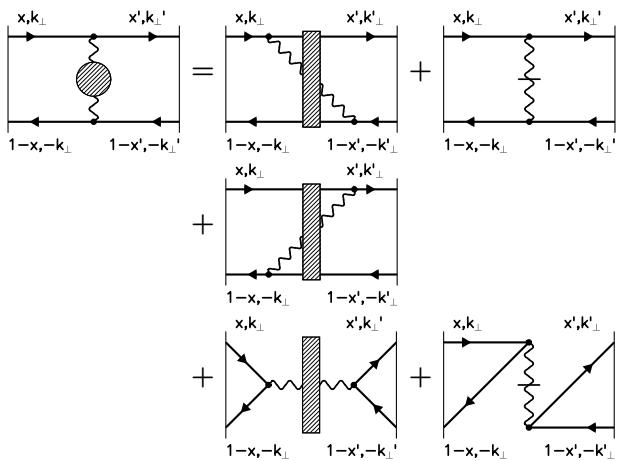

\begin{tabular}{ccccccc}
\psfig{figure=\graphpath all_effective.epsi,width=2.3cm,angle=-90}
&\hspace{0.0cm}\raisebox{1.0cm}[-1.4cm]{\bf =}&  
\psfig{figure=\graphpath dyn_effective.epsi,width=2.3cm,angle=-90} 
&\raisebox{1.0cm}[-1.4cm]{\bf +} 
&\psfig{figure=\graphpath seagull.epsi,width=2.3cm,angle=-90} \\
&\raisebox{1.0cm}[-1.4cm]{\bf +}&
\psfig{figure=\graphpath dyn2_effective.epsi,width=2.3cm,angle=-90}
& &\\
&\raisebox{1.0cm}[-1.4cm]{\bf +}&
\psfig{figure=\graphpath anni_effective.epsi,width=2.3cm,angle=-90}
&\raisebox{1.0cm}[-1.4cm]{\bf +}
& \psfig{figure=\graphpath anni_seag.epsi,width=2.3cm,angle=-90}
\end{tabular}
\vspace{0.3cm}
\caption{\label{PosiEff}
The graphs of the positronium model. Effective photon lines are 
labelled by hashed rectangles. The graphs of the annihilation interaction are
at the bottom line.}
\end{figure}
\end{minipage}

\vspace*{0.3cm}
Although we projected the higher Fock sectors onto the lower 
ones to construct the effective interaction up to now, one is, 
of course, free to project the (lower) 
$|\gamma\rangle$-sector
onto the (higher) $|e\bar{e}\rangle$-sector,
and obtains 
\begin{eqnarray}
H_{\rm eff}(\omega)= H_{PP} 
&+& H_{PN}\frac{1}{\omega-H_{NN}}H_{NP}\nonumber\\
&+& H_{PQ}\frac{1}{\omega-H_{QQ}}H_{QP} 
\end{eqnarray}

Of course we do this for convenience; we could just as well solve the 
eigenvalue problem of Eq.~(\ref{NPSpace}).
The projection is depicted in the two graphs in the bottom line of 
Fig.~\ref{PosiEff}. One is the {\em dynamic} annihilation
graph, the other is the corresponding {\em seagull}
annihilation graph. The latter is a $P$-space graph
and was omitted in Ref.~\cite{Trittmann97a} following the 
gauge principle of {\sc Tang} 
\cite{TangBrodskyPauli91}.

\section{Spectrum including the annihilation channel}
\label{Section:Spectrum}

The spectrum including the annihilation channel shows the 
expected properties: the singlet eigenvalues remain unchanged, while 
the triplet states change. 
An essential point in the actual calculations is that one has to use the
same counterterms for the Coulomb singularity as used {\em without}
the annihilation channel. 
This is due to the fact that the one photon annihilation part of the
interaction has no additional singularity 
that needs to be taken care of numerically, because of the simple
energy denominator, Eq.~(\ref{Energienenner}).
We compiled our results in the form of binding coefficients
in Table \ref{Tablespecanni}. 

\begin{minipage}{8.6cm}
\begin{figure}
\psfig{figure=\graphpath diff_anni_J1.epsi,width=8.6cm}
\vspace{0.3cm}
\caption{\label{diffanni}
Deviation of corresponding eigenvalues for $J_z{=}0$ 
and $J_z{=}1$ in\-cluding the annihilation channel
as a function of the number of
integration points $N$ for $\alpha{=}0.3$, $\Lambda{=}1.0\, m_f$.
The graph shows $\Delta M^2{:=}$ $M^2_n(J_z=0)-M^2_n(J_z=1)$
for the states $1^3S_1\,(\triangle)$, $2^3P_1\,(\Box)$, and $2^1P_1\,(\circ)$.} 
\end{figure} 
\end{minipage}

One notes a slightly larger breaking of rotational symmetry as in the case 
of missing annihilation channel. 
The triplet ground states
of different $J_z$ are still degenerate to a very good  approximation, 
but the discrepancy is bigger than without the annihilation channel.
The dependence of these discrepancies between corresponding eigenvalues for 
$J_z{=}0$ and $J_z{=}1$ on the number of integration points is shown in 
Fig.~\ref{diffanni}. The behavior of the curves is similar to those 
of the calculations without annihilation channel, {\em cf.}~Fig.~13
of Ref.~\cite{Trittmann97a}.  
An additional plot, Fig.~\ref{diffanni}, shows the dependence 
of the rotational symmetry breaking on the cutoff $\Lambda$. The 
discrepancies of corresponding eigenvalues are almost independent of
the cutoff $\Lambda$.


\begin{figure}[t]
\centerline{\psfig{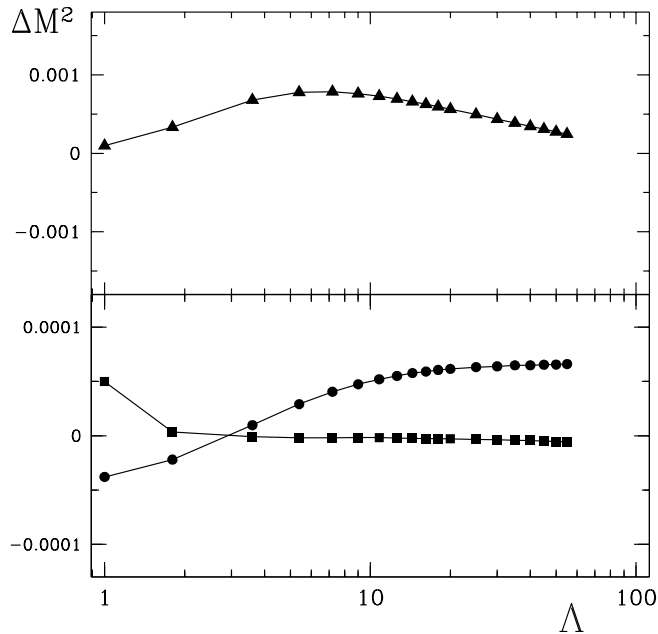}}
\vspace{0.3cm}
\caption{\label{difflambda}
Deviation of corresponding eigenvalues for $J_z{=}0$ and $J_z{=}1$ including
the annihilation channel
as a function of the cutoff $\Lambda$ for $\alpha{=}0.3, \Lambda{=}1.0\, m_f$.  
The graph shows 
$\Delta M^2:= M^2_n(J_z{=}0)-M^2_n(J_z{=}1)$
for the states $1^3S_1\,(\triangle)$, $2^3P_1\,(\Box)$, and $2^1P_1\,(\circ)$.} 
\end{figure} 

To make a comparison of our results to those of perturbation theory easier,
we show in Fig.~\ref{yrastn2} the values for the principal quantum number 
$n{=}2$ for both 
theories graphically. The structure of the two 
plots is almost the same, 
only the $2^1S_0$ state and the $2^3P_0$ state are exchanged in our results. 
This is due to the 
cutoff dependence of the S-state, which is larger than that of the other 
states (cf.~Fig.~\ref{lambdaanni}(b) and next paragraph).       
We stress that the results of the perturbative calculations change considerably,
when the
next higher order in $\alpha$ is considered. For example, the mass squared
of the triplet $2^3S_1$ is, according to {\sc Fulton} and  {\sc Martin} 
\cite{Fulton54},
$M^2(2^3S_1)=3.9780186070$ up to order $\alpha^3 {Ry\;}$, 
which is nearer to our result than the value of 
perturbative calculations up to order ${\cal O}(\alpha^4)$, displayed
in Fig.~\ref{yrastn2}. 
\noindent
\begin{minipage}{8.6cm}
\begin{table}
\begin{tabular}{r|ccc}
\rule[-3mm]{0mm}{8mm}$\mbox{Cutoff: }\Lambda$ & $B_s$ & $B_t$ & $C_{hf}$\\
\hline\hline 
   1.8\hspace{0.5cm} &  1.16373904 & 0.96234775 &  0.55942025 \\
   3.6\hspace{0.5cm} &  1.25570163 & 0.96446614 &  0.80898748 \\  
   5.4\hspace{0.5cm} &  1.29978067 & 0.96482118 &  0.93044303 \\  
   7.2\hspace{0.5cm} &  1.32941926 & 0.96541695 &  1.01111752 \\  
   9.0\hspace{0.5cm} &  1.35224000 & 0.96603457 &  1.07279285 \\  
  10.8\hspace{0.5cm} &  1.37112216 & 0.96661006 &  1.12364471 \\ 
  12.6\hspace{0.5cm} &  1.38744792 & 0.96713137 &  1.16754595 \\  
  14.4\hspace{0.5cm} &  1.40198469 & 0.96760110 &  1.20662108 \\  
  16.2\hspace{0.5cm} &  1.41520247 & 0.96802548 &  1.24215830 \\  
  18.0\hspace{0.5cm} &  1.42740143 & 0.96841025 &  1.27497551 \\ \hline 
\hline ETPT\hspace{0.3cm} & 1.11812500 & 0.90812500 & 0.58333333\\ 
${\cal O}(\alpha^6\ln\alpha)$ & \multicolumn{2}{c}{  }& 0.48792985\\ 
\end{tabular}
\vspace{0.3cm} 
\caption{\label{TableLambdaAnni}
The binding coefficients 
of the singlet ($B_s$), the triplet ($B_t$), 
and the hyperfine coefficient $C_{hf}$ are listed 
for $\alpha=0.3,$ $N_1=41, N_2=11$. For a comparison, the coefficients of 
equal time perturbation theory are also listed, up to order
${\cal O}(\alpha^4)$ (ETPT) and to ${\cal O}(\alpha^6\ln\alpha)$}
\end{table}
\end{minipage}

\section{Parameter dependence of the spectrum}
\label{Section:ParamDep}

The convergence of the eigenvalues with growing number of integration 
points $N$ is the same as the case of no annihilation channel. 
To be explicit, the convergence of the eigenvalues can be shown to be
exponential. We performed a $\chi^2$ fit to the function
\begin{equation}
f(N)=a-b\exp\left\{(N-N_0)/c\right\}.
\end{equation}
and obtain for the singlet ground state $1\,^{1\!}S_0$
$a=  3.9061000 \pm  0.0000008$,
$b= (9.6399 \pm  0.0193)\times 10^{-4},$ and
$c= 2.2372 \pm 0.0063,$
with $\chi^2=5.7$.
We did not perform the limit $N\rightarrow\infty$, because the accuracy of the 
results for $N>20$ suffices to compare to other data. 

\begin{figure}[t]
\centerline{\psfig{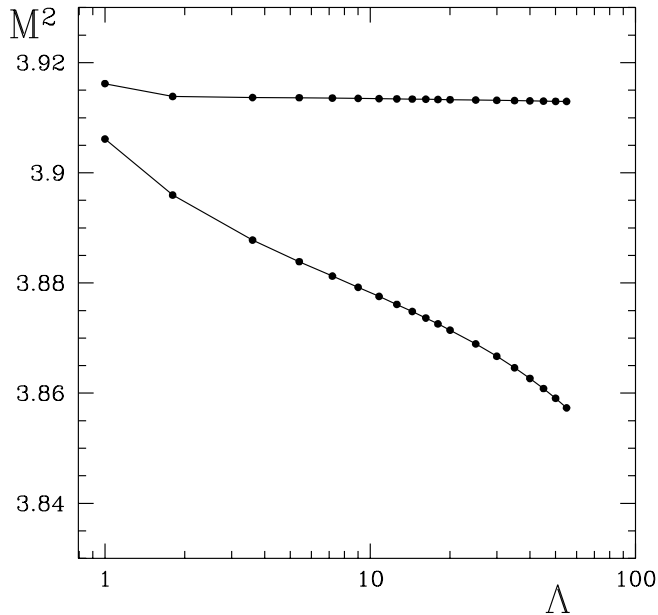}}
\vspace{0.3cm}
\caption{\label{lambdaanni}
The spectrum with 
annihilation channel as a function of the cutoff $\Lambda$: 
ground states ($n{=}1$).
The parameters for the calculation are
$\alpha=0.3$, $J_z=0$, $N_1=25$, $N_2=21$.}
\end{figure}

The cutoff dependence of the positronium spectrum including the 
annihilation channel is comparable to that of the spectrum without it. 
However, a striking difference occurs: the inclusion of the annihilation
channel stabilizes the cutoff dependence of the eigenstates. 
In particular, the triplet ground state in Fig.~\ref{lambdaanni}(a)
shows only a small dependence on the cutoff, when one compares it to the 
behavior of the same state in the calculations without annihilation channel
\cite{Trittmann97a}.
One can fit these curves with a polynomial in $\log \Lambda$. The
singlet ground state eigenvalues are the same as without the annihilation graph
and for the triplet one obtains
\begin{equation}
M^2_{t}(\Lambda)\simeq 3.91392-0.00029 \log\Lambda
                                  +0.00015 \log^2\Lambda.
\end{equation}
Comparison with Eq.~(22) of Ref.~\cite{Trittmann97a} shows that 
the decrease of the triplet 
with $\log\Lambda$ is suppressed by including the annihilation channel by
a factor of 60.
Also the excited states, $n{=}2$, show a different behavior, when compared 
to Fig.~13 of Ref.~\cite{Trittmann97a}. 
Here, only one state shows level crossing as
$\Lambda$ grows large. The eigenvalue of the $2^3P_1$ state, however, 
depends only weakly on the cutoff. 


\begin{figure}[t]
\centerline{
\psfig{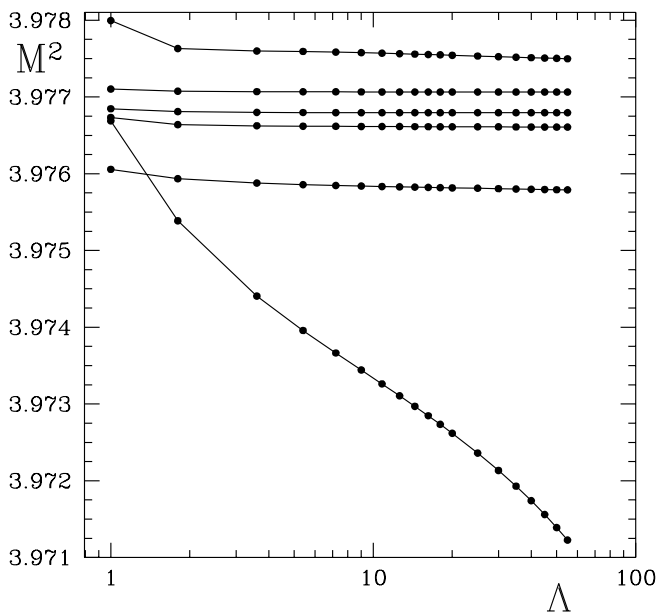}
}
\vspace{0.3cm}
\caption{\label{lambdaanni2}
The spectrum with 
annihilation channel as a function of the cutoff $\Lambda$: 
first radial excited states ($n{=}2$).
The parameters for the calculation are
$\alpha=0.3$, $J_z=0$, $N_1=25$, $N_2=21$.
}
\end{figure} 

The values for the binding coefficients $B_n$ and the coefficient of the 
hyperfine splitting $C_{hf}$ are presented in Table~\ref{TableLambdaAnni}.
The values are correct for 
a cutoff $\Lambda\simeq 2 m_f$ when compared to results of 
perturbation theory up to ${\cal O}(\alpha^4)$. However, 
the effects of higher order correction to perturbative calculations
are significant
for a large coupling such as $\alpha=0.3$. 
The result of perturbation theory to ${\cal O}(\alpha^6\ln \alpha)$
for the coefficient $C_{hf}$ is considerably smaller than that to order 
${\cal O}(\alpha^4)$.  

Concluding, we state that the cutoff dependence of the spectrum 
is improved as compared with the case of missing annihilation
channel. This makes it evident that the annihilation channel is a necessary
part of the theory. 
Note that in muonium, the annihilation state is absent. This can lead to
problems in interpreting the stabilizing feature of the channel 
\cite{Brodsky97}.

As a further investigation of the properties of our model, we will vary the
coupling constant and interpret the spectrum. A similar procedure
was performed by {\sc Dyks\-hoorn et al.} \cite{Dykshoorn90a,Dykshoorn90b}, 
who studied 
coupled integral equations for QED-bound states in equal-time quantization
with a variational ansatz. They calculate masses for the 
lowest eigenstates of positronium with and without the annihilation channel
and plot them versus the coupling constant. The prominent feature of their
figures is the occurrence of a critical coupling at which the masses become 
smaller than zero. 
We have performed the analogous calculations within our approach. 
It seems at first glance [Fig.~\ref{alphaspectrum}(a)] as if the eigenvalues,
after decreasing 
quadratically with the coupling as expected by perturbation theory,
stabilize at a coupling $\alpha\simeq 1.5$.
However, further investigation shows that this 
is merely an effect of the cutoff dependence of the spectrum. The eigenvalues
in  Fig.~\ref{alphaspectrum}(a) were calculated 
with a cutoff $\Lambda=1.0\; m_f$, which
is too small for a coupling constant of $\alpha=1.0$, since then the Bohr
momentum is of the same order as the cutoff.
Fig.~\ref{alphaspectrum}(b) shows clearly that there is a critical coupling.
The masses calculated with a cutoff of $\Lambda=20\, m_f$ tend to zero 
at $\alpha\simeq 0.5$. A similar value was found in \cite{Dykshoorn90b}.

\section{Discussion}
\label{Section:Discussion}


In the present work we completed in a sense the theory of positronium 
in {\em front form} dynamics set up in Ref.~\cite{Trittmann97a}.  
The theory was enlarged by including the one photon annihilation channel.
The inclusion is a test for the consistency
of the model. In particular, the conception of an effective 
theory operating with resolvents can 
be falsified if the annihilation channel cannot be implemented without
special assumptions.
Our results show that the implementation of the annihilation channel
is unproblematic in the sense that the same formalism can be applied as 
in the case of the projection of the effective $|e\bar{e}\gamma\rangle$-sector
onto the electron-positron sector.
As an interesting
property of the annihilation channel we find a strict separation
of the instantaneous and the non-instantaneous interaction:
the seagull interaction is present only in the $J_z=0$ sector, whereas the
dynamic graph has non-vanishing matrix elements only for $J_z=\pm 1$.
Our approach passes another test: both graphs 
yield the same contributions to the eigenvalues and consequently 
rotational symmetry is seen also in the
spectrum including the annihilation channel.
Moreover, the inclusion of the annihilation channel 
improves the results for the hyperfine splitting. 

We stress the point that the implementation of the annihilation channel
completes the investigations of how to construct an effective 
interaction for the electromagnetic fermion-antifermion system in the meaning
of the method of iterated resolvents. 
We have put all effects of higher Fock states
into an effective $|e\bar{e}\gamma\rangle$-sector, as far as the 
{\em spectator interaction} is concerned, {\em i.e.}~the 
interactions in which the photons
are not directly involved. The remainder of the interaction relies
in the coupling function of the vertices. 
A hint to this conclusion is the logarithmic cutoff dependence 
of our results. It is clear that the coupling constant depends on the cutoff
and has to be analyzed.
A future aim, beyond the scope of this work, will be to show that 
the physical results of our model 
become independent of the cutoff, as soon as 
renormalization group techniques are consistently applied.   
This is supported by the fact that we find a stabilizing effect of the
annihilation channel on the dependence of the spectrum on the cutoff:
all eigenvalues show slower variation with growing cutoff $\Lambda$ when
the annihilation graph is added,
in some cases this even prevents level crossings with other states.
However, in muonium for instance, this channel is absent and it remains unclear
which interpretation this should yield.
We therefore conclude that our model is correct as long as the  
vacuum polarization effects are not considered.
 
The possibility of one boson annihilation is the main difference
between QED and QCD in effectively truncated Fock spaces 
where the dynamic three- or
four-gluon interaction is not possible but resides in the coupling function. 
The one boson sector exists only in QED because of the constraint of 
color neutrality in QCD. 
We showed that this sector can be projected onto the (already) effective  
electron-positron sector. 

%
\noindent
\begin{minipage}{8.6cm}
\begin{table}
\begin{tabular}{rc|l|l|r}
\rule[-3mm]{0mm}{8mm}$n$ & Term & $\hfill B_n(J_z{=}0)\hfill $ & $\hfill B_n(J_z{=}+1)\hfill$ & $\hfill \Delta B_n \hfill$ \\ \hline \hline
  1&$1^1S_0$ & 1.049552(17) & $\hfill\mbox{---}\hfill$ & $\hfill\mbox{---}\hfill$ \\ \hline 
  2&$1^3S_1$ & 0.936800(189) & 0.937902(151) & -0.001102 \\ \hline 
\hline
  3&$2^1S_0$ & 0.260237(169) & $\hfill\mbox{---}\hfill$ & $\hfill\mbox{---}\hfill$ \\ \hline 
  4&$2^3S_1$ & 0.255292(184) & 0.255359(179) & -0.000067 \\ \hline 
  5&$2^1P_1$ & 0.257969(160) & 0.258037(168) & -0.000068 \\ \hline 
  6&$2^1P_1$ & 0.267090(160) & $\hfill\mbox{---}\hfill$ & $\hfill\mbox{---}\hfill$ \\ \hline 
  7&$2^3P_1$ & 0.259667(206) & 0.260013(163) & -0.000346 \\ \hline 
  8&$2^3P_2$ & 0.245615(239) & 0.245755(228) & -0.000140 \\ \hline 
\hline
  9&$3^1S_0$ & 0.115206(314) & $\hfill\mbox{---}\hfill$ & $\hfill\mbox{---}\hfill$ \\ \hline 
 10&$3^3S_1$ & 0.113434(156) & 0.113497(179) & -0.000063 \\ \hline 
 11&$3^1P_1$ & 0.114490(269) & 0.114521(283) & -0.000032 \\ \hline 
 12&$3^1P_1$ & 0.117142(272) & $\hfill\mbox{---}\hfill$ & $\hfill\mbox{---}\hfill$ \\ \hline 
 13&$3^3P_1$ & 0.115127(326) & 0.115102(275) & 0.000025 \\ \hline 
 14&$3^3P_2$ & 0.113731(284) & 0.113753(283) & -0.000023 \\ \hline 
 15&$3^1D_2$ & 0.112816(150) & 0.112842(158) & -0.000027 \\ \hline 
 16&$3^3D_1$ & 0.112977(161) & 0.112987(164) & -0.000010 \\ \hline 
 17&$3^3D_2$ & 0.112520(158) & 0.112524(160) & -0.000004 \\ \hline 
 18&$3^3D_3$ & 0.111027(377) & 0.111072(373) & -0.000045 \\ \hline 
\hline
 19&$4^1S_0$ & 0.065490(588) & $\hfill\mbox{---}\hfill$ & $\hfill\mbox{---}\hfill$ \\ \hline 
 20&$4^3S_1$ & 0.064707(480) & 0.064723(481) & -0.000016 \\ \hline 
 21&$4^1P_1$ & 0.065003(467) & 0.065061(486) & -0.000059 \\ \hline 
 22&$4^1P_1$ & 0.066119(470) & $\hfill\mbox{---}\hfill$ & $\hfill\mbox{---}\hfill$ \\ \hline 
 23&$4^3P_1$ & 0.065331(487) & 0.065276(475) & 0.000055 \\ \hline 
 24&$4^3P_2$ & 0.064265(309) & 0.064372(348) & -0.000107 \\ \hline 
 25&$4^1D_2$ & 0.063968(430) & 0.064041(363) & -0.000073 \\ \hline 
 26&$4^3D_1$ & 0.064099(319) & 0.064090(325) & 0.000009 \\ \hline 
 27&$4^3D_2$ & 0.063870(391) & 0.063881(477) & -0.000011 \\ \hline 
 28&$4^3D_3$ & 0.063788(387) & 0.063807(381) & -0.000019 \\ \hline 
 29&$4^1F_3$ & 0.063141(96) & 0.063112(104) & 0.000030 \\ \hline 
 30&$4^3F_2$ & 0.063299(112) & 0.063233(116) & 0.000065 \\ \hline 
 31&$4^3F_3$ & 0.063234(119) & 0.063209(119) & 0.000024 \\ \hline 
 32&$4^3F_4$ & 0.063103(104) & 0.063422(158) & -0.000319 \\ \hline 
\hline
 33&$5^1S_0$ & 0.043253(806) & $\hfill\mbox{---}\hfill$ & $\hfill\mbox{---}\hfill$ \\ \hline 
 34&$5^3S_1$ & 0.043046(682) & 0.042912(739) & 0.000134 \\ \hline 
 35&$5^1P_1$ & 0.043412(1100) & 0.042724(853) & 0.000688 \\ 
\end{tabular}
\vspace{0.3cm}
\caption{\label{Tablespecanni}
The positronium spectrum for $\alpha=0.3,$ $\Lambda=1.0\, m_f, N_1=N_2=21$.
The non-relativistic notation for the terms and the binding coefficients
$B_n:=(4-M_n)/\alpha^2$ for $J_z{=}0$ and $J_z{=}+1$ 
including the annihilation channel are shown. The discrepancy between the eigenvalues is $\Delta B_n:=B_n(J_z{=}0)-B_n(J_z{=}+1)$. The numerical errors are estimated from the difference between the values for maximum and next to maximum number of integration points. The actual errors are smaller due to 
the exponential convergence of the eigenvalues with $N$.
The $k$ numbers in brackets are the
 errors in the last $k$ digits.}
\end{table}
\end{minipage}

\begin{figure}[h]
\centerline{
\psfig{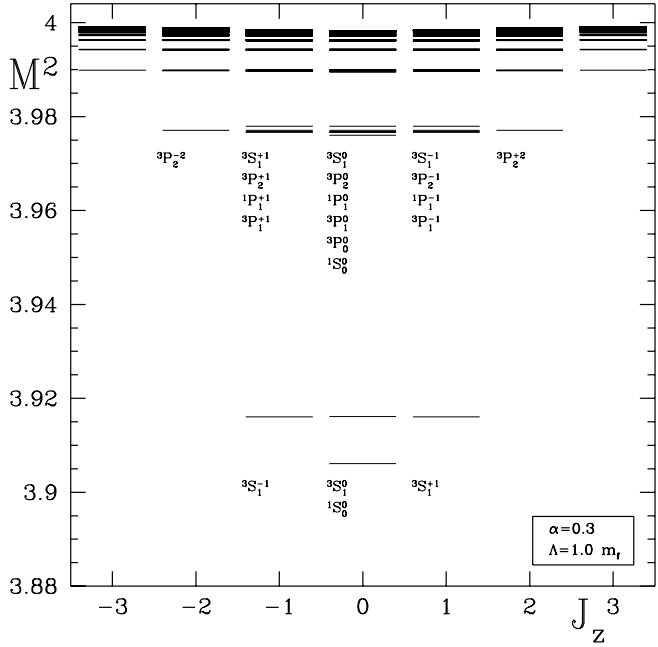}}
\vspace{0.3cm}
\caption{\label{yrastanni}
Compiled spectra of 
positronium with different $J_z=-3,-2,\ldots,+3$
including the annihilation channel.
All spectra have been calculated with $\alpha=0.3$, 
$\Lambda=1.0\, m_f$, $N_1=N_2=21$.
The mass squared eigenvalues $M^2_n$ in units of the electron mass $m^2_f$
are shown.  
The notation for the states is $^{3S+1}L^{J_z}_J$.
The resolution of the plot is inadequate for the 
multiplets. Nevertheless, the numerical degeneracy of the three triplet
ground states $^3S^{-1}_1$,$^3S^0_1$, and $^3S^1_1$ becomes very clear.
} 
\end{figure}
\begin{figure}
\centerline{\psfig{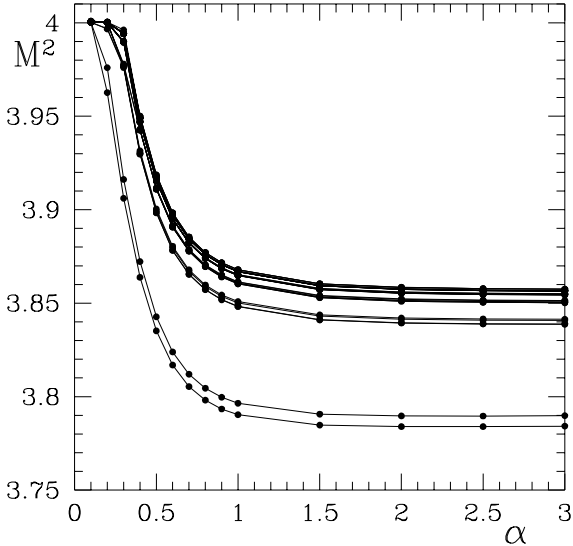}}
\vspace{0.3cm}
\caption{\label{alphaspectrum}
The spectrum as a function of the coupling 
constant for $\Lambda=1.0\,m_f$. Here, the  
eigenvalues seem to converge to a stable value as $\alpha$ grows large.
However, this is just an effect of the cutoff dependence of the spectrum:
for a larger cutoff $\Lambda=20\,m_f$ the eigenvalues become negative
at a critical coupling $\alpha_c\simeq 0.5$.}
\end{figure} 

\begin{figure}
\centerline{\psfig{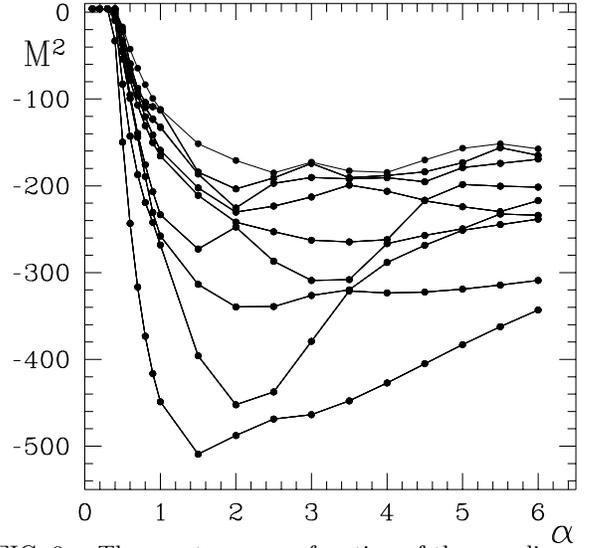}}
\caption{\label{alphaspectrum2}
The spectrum as a function of the coupling constant for 
$\Lambda=20.0\,m_f$. For this large cutoff the eigenvalues become negative
at a critical coupling $\alpha_c\simeq 0.5$.}
\end{figure} 

\begin{appendix}

\section{Calculation of effective matrix elements}
\label{Appx:MatrixElements}

The implementation of the annihilation channel is straightforward, 
but non-trivial.
We describe shortly the calculation 
in the $(J_z{=}0)$ and $(J_z{=}\pm 1)$ sectors. 
The corresponding diagrams, displayed in the lowest line of 
Fig.~\ref{PosiEff},
vanish in all
other sectors because of the helicity of the photon: no angular momentum larger
than $J=1$ is possible.
The functions derived depend on the light-cone momenta, $(x,\,\vec{k}_{\perp})$,
and are given in Table \ref{HelicityTableAnnihilation}. They have to be added 
to the functions listed in Ref.~\cite{Trittmann97a}, Table V, to obtain
the full Hamiltonian. 
Following the procedure described in Ref.~\cite{Trittmann97a} 
({\em cf.} also Ref.~\cite{TrittmannPauli97}), we use for the 
energy denominator the following averaged kinetic energy
\begin{equation}\label{OmegaStar}
T^*(x,\vec{k}_{\perp};x',\vec{k}'_{\perp})
        =\frac{1}{2}\left( \frac{m_f^2 + 
        \vec{k}_{\perp}}{x(1-x)} + \frac{m_f^2 + \vec{k}'_{\perp}}{x'(1-x')}
        \right).
\end{equation}
The denominator in the one photon sector is simple because the 
photon has zero mass:
\begin{equation}\label{Energienenner}
G(\omega) = \frac{1}{\omega - H_{\gamma}}= \frac{1}{T^*}.
\end{equation}
Note that this denominator does {\em not} depend on the directions of the 
vectors $\vec{k}_{\perp}$, $\vec{k}_{\perp}'$, {\em i.e.}~on the angles $\varphi$, $\varphi'$.

The matrix elements of the vertex interaction $V_{g\rightarrow q\bar{q}}$ 
({\em cf.~e.g.} Ref.~\cite{Tang91}) 
split up into their three components with
different helicity factors, are
\begin{eqnarray}
\langle e\bar{e}|V_A|\gamma\rangle&=&-m\sqrt{\beta}
\frac{1}{x(1-x)}
\times\delta^{+\lambda_1}_{+\lambda_2}\delta^{+\lambda_1}_{+\lambda_3},
\nonumber\\
\langle e\bar{e}|V_B|\gamma\rangle&=&-\sqrt{\beta}
\frac{k_{\perp}}{1-x}e^{+i\lambda_2\varphi}
\times\delta^{+\lambda_1}_{+\lambda_2}\delta^{+\lambda_1}_{-\lambda_3},
\nonumber\\
\langle e\bar{e}|V_C|\gamma\rangle&=&\sqrt{\beta}
\frac{k_{\perp}}{x}e^{-i\lambda_2\varphi}
\times\delta^{+\lambda_1}_{-\lambda_2}\delta^{+\lambda_1}_{+\lambda_3}.
\end{eqnarray}
The complete matrix elements are obtained by symmetry, since
$
\langle e\bar{e}|V_i|\gamma\rangle=
\overline{\langle \gamma|V_i|e\bar{e}\rangle}.
$
To classify the states according to their quantum numbers with respect
to the third (kinetic) component of the total angular momentum, ${\cal J}_3$,
we project on sectors of definite $J_z$.
It turns out that because of the simple energy denominator, 
Eq.~(\ref{Energienenner}), the only dependence on the angles comes from the
projectors on the $J_z$ sectors and is proportional to 
$\cos\{\varphi-\varphi'\}$
or $\sin\{\varphi-\varphi'\}$. As a result, all 
matrix elements of the {\em dynamic} annihilation graph for $J_z{=}0$ 
vanish when integrated over the angles.
Only for $J_z{=}\pm 1$ do some of the matrix elements survive the integration. 

Because of the combination of matrix elements with the projectors onto the 
$J_z$ sectors,
two types of functions emerge for $J_z{=}\pm1$:
one is independent of the angles, the other has a dependence 
proportional to $\exp\{\pm 2i(\varphi-\varphi')\}$.
The latter vanishes after angular integration. The helicity table 
\ref{SymbolicHelicityTable} is given to illustrate the helicity dependencies.
It holds for $J_z{=}+1$. The analogous table for $J_z{=}-1$ is obtained 
by the operation 
\[
W_{ij}(J_z{=}\!+\!1;\,\lambda_1,\lambda_2)=-\lambda_1
W_{ij}(J_z{=}\!-\!1;\,-\lambda_1,-\lambda_2).
\]

The simple kinematics ($x_e+x_{\bar{e}} = 1$) of the
seagull annihilation graph 
result in a constant contribution of this graph to the Hamiltonian 
matrix. 
It is 
\begin{equation}
\langle e\bar{e}|S|e\bar{e}\rangle = -2\beta
\delta^{+\lambda_2}_{-\lambda_1}\times\delta^{+\lambda_2'}_{-\lambda_1'}.
\end{equation}
Because of its helicity factors, the graph acts 
only between states with 
\begin{equation}\label{SpinBed}
S_z = S_z' = 0.
\end{equation}  
This means that the seagull graph does not contribute when $J_z{\neq}0$, because
it has a factor proportional to $\cos(\varphi-\varphi')$ resulting from 
(\ref{SpinBed}). 
A rather surprising consequence is that the dynamic graph contributes only 
for $J_z=\pm 1$, the instantaneous graph only for $J_z=0$. 
However, to maintain rotational invariance, both diagrams must 
yield the same value, though one shows much more structure than the other. 
Degeneracy of the orthopositronium ground state
with respect to $J_z$ 
is found without the annihilation channel and inclusion must not destroy it.
At first glance, there seems to be a manifest breaking of rotational symmetry:
the helicity table \ref{SymbolicHelicityTable} separates between states
with $(\lambda_1,\lambda_2)=(\uparrow\uparrow)$ and
$(\lambda_1,\lambda_2)=(\downarrow\downarrow)$. But this is only a consequence
of the integration over the angles: for $J_z=+1$ the $(\downarrow\downarrow)$-
combination
gives no contribution, and likewise does the $(\uparrow\uparrow)$-combination 
for $J_z=-1$.

\noindent
\begin{minipage}{8.6cm}
\begin{table}
\begin{tabular}{|c||c|c|c|c|}\hline
\rule[-3mm]{0mm}{8mm}{\bf out : in} & $\lambda_1',\lambda_2'=
\uparrow\uparrow$ 
& $\lambda_1',\lambda_2'=\uparrow\downarrow$ 
& $\lambda_1',\lambda_2'=\downarrow\uparrow$
& $\lambda_1',\lambda_2'=\downarrow\downarrow$ \\ \hline\hline
\rule[-3mm]{0mm}{8mm}$\lambda_1,\lambda_2=\uparrow\uparrow$ & 
$W_{AA}$   
&$W_{AB}$ & $W_{AC}$ & $0$ \\ \hline
\rule[-3mm]{0mm}{8mm}$\lambda_1,\lambda_2=\uparrow\downarrow$ & 
$W_{BA}$ 
& $W_{BB}$ & $W_{BC}$ & $0$ \\ \hline
\rule[-3mm]{0mm}{8mm}$\lambda_1,\lambda_2=\downarrow\uparrow$& 
$W_{CA}$ & $W_{CB}$ & $W_{CC}$  & $0$\\ \hline
\rule[-3mm]{0mm}{8mm}$\lambda_1,\lambda_2=\downarrow\downarrow$ & $0$ 
& $0$ & $0$ & $0$  \\
\hline
\end{tabular}
\vspace{0.3cm}
\caption{\label{SymbolicHelicityTable}
Symbolic helicity table for the dynamic annihilation graph.
The functions $W_{ii}$ are identical with the expressions $F_i$ listed in Table 
\protect\ref{HelicityTableAnnihilation}. Here, terms proportional to 
$\delta_{|J_z|,0}$ are omitted.}
\end{table}
\end{minipage}
\noindent


The dependence of the effective interaction on the helicities of 
in- and out-going particles is displayed in the form of tables.
The following notation is used for functions of the type
$F(x,k_{\perp};x',k_{\perp}')$.
An asterisk denotes the permutation of particle and anti-particle
\[
F_3^*(x,k_{\perp};x',k_{\perp}'):=F_3(1-x,-k_{\perp};1-x',-k_{\perp}').
\]
If the function additionally depends 
on the component of the total angular momentum $J_z=n$,
a tilde symbolizes the operation
$\tilde{F}_i(n)=F_i(-n)$.

\subsection{Helicity table of the annihilation graph}
\label{AnniTabelle}

\noindent
The functions $F_i(1;2):=F_{i}(x,k_{\perp};x',k_{\perp}')$ displayed in 
Table \ref{HelicityTableAnnihilation} are: 
\begin{eqnarray*}
F_1(x,k_{\perp};x',k_{\perp}') &:=& \frac{\alpha}{\pi}\frac{2}{\omega^*}
         \frac{m^2}{xx'(1-x)(1-x')}\delta_{|J_z|,1}\\
F_2(x,k_{\perp};x',k_{\perp}') &:=&\frac{\alpha}{\pi}\left(\frac{2}{\omega^*}
         \frac{k_{\perp}k_{\perp}'}{xx'}
         \delta_{|J_z|,1}+4\delta_{J_z,0}\right)\\
F_3(x,k_{\perp};x',k_{\perp}') &:=& \frac{\alpha}{\pi}\frac{2}{\omega^*}
	 \lambda_1
         \frac{m}{x(1-x)}
         \frac{k_{\perp}'}{1-x'}\delta_{|J_z|,1}\\
F_4(x,k_{\perp};x',k_{\perp}') &:=& -\frac{\alpha}{\pi}\left(\frac{2}{\omega^*}
        \frac{k_{\perp}k_{\perp}'}{x'(1-x)}\delta_{|J_z|,1}-4\delta_{J_z,0}
	\right).
\end{eqnarray*}
The table for $J_z=-1$ is obtained by inverting {\em all} helicities.
Note that the table has non-vanishing matrix elements
for $|J_z|\leq 1$ only.
This restriction is due to the angular momentum of the photon.

\noindent
\begin{minipage}{8.6cm}
\begin{table}[h]
\begin{tabular}{|c||c|c|c|c|}\hline
\rule[-3mm]{0mm}{8mm}{\bf out:in} & $\lambda'_1\lambda'_2=
\uparrow\uparrow$
& $\lambda'_1\lambda'_2=\uparrow\downarrow$ 
& $\lambda'_1\lambda'_2=\downarrow\uparrow$ &
$(\lambda'_1,\lambda'_2=\downarrow\downarrow$ \\ \hline\hline
\rule[-3mm]{0mm}{8mm}$\lambda_1\lambda_2=\uparrow\uparrow$ & 
$F_1(1;2)$   
&$F_3(2;1)$ & $F^*_3(2;1)$ & $0$ \\ \hline
\rule[-3mm]{0mm}{8mm}$\lambda_1\lambda_2=\uparrow\downarrow$ & 
$F_3(1;2)$ 
& $F^*_2(1;2)$ & $F_4(2;1)$ &$0$ \\ \hline
\rule[-3mm]{0mm}{8mm}$\lambda_1\lambda_2=\downarrow\uparrow$& 
$F_3^*(1;2)$ & $F_4(1;2)$ & $F_2(1;2)$  & $0$\\ \hline
\rule[-3mm]{0mm}{8mm} $\lambda_1\lambda_2=\downarrow\downarrow$ & $0$ 
& $0$ & $0$ & $0$  \\
\hline
\end{tabular}
\vspace{0.3cm}
\caption{\label{HelicityTableAnnihilation}
Helicity table of the annihilation graph for $J_z\ge 0$.}
\end{table}
\end{minipage}

\end{appendix}



\bibliography{Paper_Anni_Biblio,\bibpath Paper_Jz_Biblio,\bibpath 
	      ParticleDetection,\bibpath Lichtkegel,\bibpath PhysikHistory}

\begin{thebibliography}{10}

\bibitem{Trittmann97a}
U. Trittmann and H.-C. Pauli, On rotational invariance in front form dynamics,
  1997, {\tt hep-th/9705021}.

\bibitem{CommentHigherStates}
Or rather the substitution of effects of higher Fock sectors with the use of
  {\em effective} matrix elements in the remaining sectors.``Higher'' here in
  the sense of ascending $n$ in Table \ref{HolyMatrixQED}.

\bibitem{Pauli96b}
H.-C. Pauli, Solving Gauge field Theory by Discretized Light-Cone Quantization,
  1996, {\tt hep-th/9608035}.

\bibitem{TangBrodskyPauli91}
A.~C. Tang, S.~J. Brodsky, and H.-C. Pauli, Phys. Rev. {\bf D} {\bf 44},  1842
  (1991).

\bibitem{Fulton54}
T. Fulton and P. Martin, Phys. Rev. {\ } {\bf 95},  811  (1954).

\bibitem{Brodsky97}
S.~J. Brodsky, private communication.

\bibitem{Dykshoorn90a}
W. Dykshoorn and R. Koniuk, Phys. Rev. {\bf A} {\bf 41},  60  (1990).

\bibitem{Dykshoorn90b}
W. Dykshoorn and R. Koniuk, Phys. Rev. {\bf A} {\bf 41},  64  (1990).

\bibitem{TrittmannPauli97}
U. Trittmann and H.-C. Pauli, Quantum electrodynamics at strong couplings,
  1997, {\tt hep-th/9704215}.

\bibitem{Tang91}
A. Tang, S.~J. Brodsky, and H.-C. Pauli, Phys. Rev. {\bf D} {\bf 44},  1842
  (1991).

\end{thebibliography}



\newpage
\noindent
\begin{figure}[h]
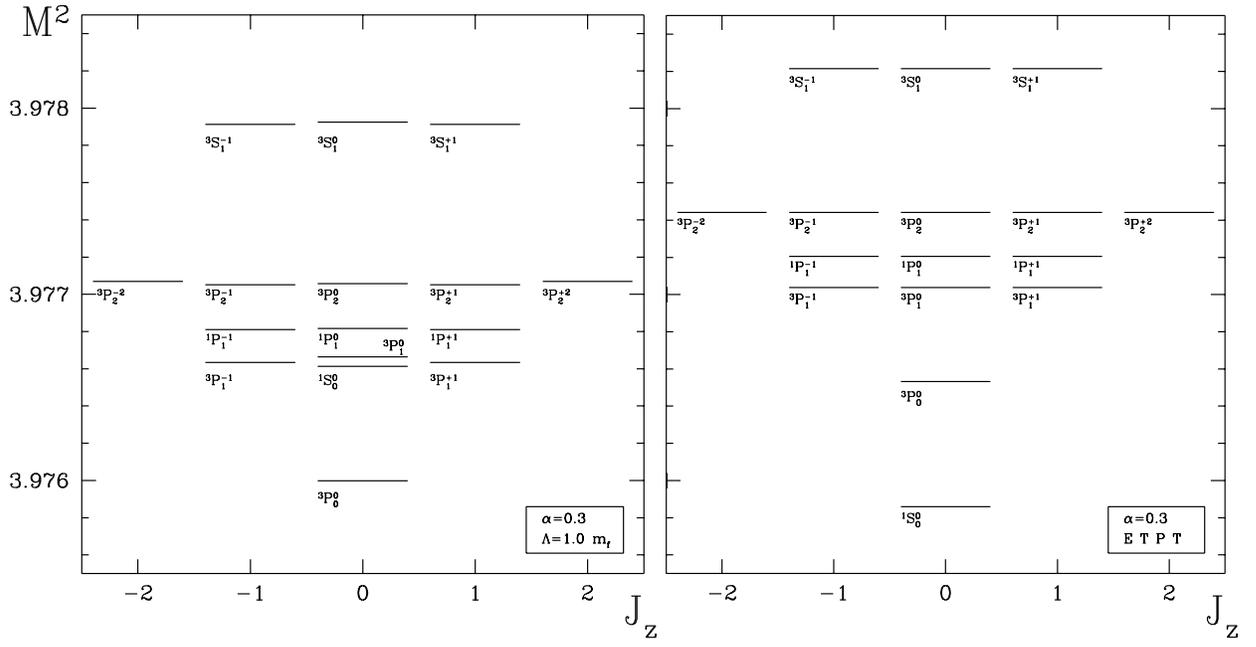

\begin{minipage}{18cm}
\centerline{
\psfig{figure=\graphpath yrast_n2.epsi,width=8.6cm,angle=0}
\psfig{figure=\graphpath yrast_n2_E.epsi,width=7.61cm,angle=0}}
\vspace{0.3cm}
\caption{\label{yrastn2}Comparison of 
multiplets for $n{=}2$: (a) results of the present work with
$\alpha=0.3$, $\Lambda= 1.0\,m_f$, $N_1=N_2=21$; 
(b) equal-time perturbation theory (ETPT) up to order ${\cal O}(\alpha^4)$.}
\end{minipage}
\end{figure}
\end{multicols}
\end{document}